\documentclass[%
aps,%
prb,%
twocolumn,
superscriptaddress,%
showpacs,%
preprintnumbers,%
byrevtex,%
floatfix%
]{revtex4}

\usepackage{ifthen}
\usepackage[usenames]{color}
\usepackage{amssymb}
\usepackage{amsfonts}
\usepackage{amsmath}
\usepackage[dvips]{graphicx}
\usepackage[dvips, colorlinks=true, bookmarks, pdftitle={A PDF file from LaTeX}, pdfauthor={MdV}, pdfsubject={From LaTeX to PDF},pdfkeywords={PDF, LaTeX, hyperlinks, hyperref}]{hyperref}

\def\ii{\textrm{i}\,\!}

\newcommand{\beann} {\begin{eqnarray*}}
\newcommand{\eeann} {\end{eqnarray*}}
\newcommand{\bea} {\begin{eqnarray}}
\newcommand{\eea} {\end{eqnarray}}
\newcommand{\labs} {\left\vert}
\newcommand{\rabs} {\right\vert}
\newcommand{\lsb} {\left[}
\newcommand{\rsb} {\right]}
\newcommand{\lrb} {\left(}
\newcommand{\rrb} {\right)}
\newcommand{\lcb} {\left\{}
\newcommand{\rcb} {\right\}}
\newcommand{\lab} {\left\langle}
\newcommand{\rab} {\right\rangle}

\newcommand{\energy} {E}
\newcommand{\ReP} {\textrm {Re}~}
\newcommand{\ImP} {\textrm {Im}~}
\newcommand{\Ham} {{\mathcal H}}
\newcommand{\Gf}  {{\mathcal G}} 
\newcommand{\Glead} {g} 
\newcommand{\sr} {{\textrm S}}
\newcommand{\la} {{\textrm L_1}}
\newcommand{\lb} {{\textrm L_2}}
\newcommand{\lc} {{\textrm L_3}}
\newcommand{\lalle} {{\textrm L_{\alpha}}}
\newcommand{\lalleb} {{\textrm L_{\beta}}}
\newcommand{\T}   {T}
\newcommand{\Cond}   {G^\textrm{tot}}
\newcommand{\CondInel}   {G^\textrm{inel}}
\newcommand{\CondEl}   {G^\textrm{el}}
\newcommand{\gc}{\,^\circ\mathrm{C}}
\newcommand{\g}{\,^\circ}

\begin{document}
\date{October 1, 2004}

\title{Defective transport properties of three-terminal carbon nanotube
junctions}

\author{Miriam Del Valle}
\affiliation{Departamento de F\'{\i}sica Te\'{o}rica de la Materia Condensada, Universidad
Aut\'{o}noma de Madrid, Facultad de Ciencias, C-V, E-28049 Madrid, Spain}
\affiliation{Institute for Theoretical Physics, University of Regensburg, D-93040 Regensburg, Germany}
\author{Carlos Tejedor}
\affiliation{Departamento de F\'{\i}sica Te\'{o}rica de la Materia Condensada, Universidad
Aut\'{o}noma de Madrid, Facultad de Ciencias, C-V, E-28049 Madrid, Spain}
\author{Gianaurelio Cuniberti}
\affiliation{Institute for Theoretical Physics, University of Regensburg, D-93040 Regensburg, Germany}

\begin{abstract}
We investigate the transport properties of three terminal carbon based nanojunctions 
within the scattering matrix approach. The stability of such junctions is subordinated 
to the presence of nonhexagonal arrangements in the molecular network. Such 
``defective" arrangements do influence the resulting quantum
transport observables, as a consequence of the possibility of acting as pinning 
centers of the correspondent wavefunction. By investigating a fairly wide class 
of junctions we have found regular mutual dependencies between such localized states 
at the carbon network and a strikingly behavior of the
conductance. In particular, we have shown that Fano resonances emerge as a natural 
result of the interference between defective states
and the extended continuum background. As a consequence, the currents through the 
junctions hitting these resonant states might experience variations on a relevant 
scale with current modulations of up to 75\%.

\end{abstract}
\pacs{%
73.22.-f, 
73.40.Sx, 
73.63.-Fg, 
81.07.De, 
85.35.Kt 
}

\maketitle
\section{Introduction}

Molecular electronics is a promising candidate in our way towards smaller electronic
devices and a lot of effort has been put, in the last decades since the
first proposal in this direction,\cite{AviramR74} into
making it a feasible goal for the near future. Nevertheless, there is still a lot of work
to be done in order to understand the behavior of nanodevices and fully
exploit their nontrivial quantum effects 
dominating the physical and chemical properties at the nanometer level.\cite{CunibertiFR05}

Our interest will focus on carbon nanotubes (CNTs), which are
basically rolled up graphitic sheets.
CNTs have been
proposed as leads and bridge molecules, since they possess a great versatility,
allowing for metallic as well as semiconducting behavior,\cite{SaitoDD98,YiPC04}
depending on their diameter and chirality, that
is, their degree of helicity. Nanotubes can be easily modified by
introducing pentagons, heptagons or octagons into their hexagonal network, as
was already shown\cite{IijimaIA92,IharaIK93} soon after their
discovery.\cite{Iijima91}
By joining a metallic nanotube with a semiconducting one, a
heterojunction is formed showing a transport behavior corresponding to
that of a rectifying diode, as already seen experimentally;\cite{YaoPBD99}
actually several applications of CNTs in
nanoscale devices have been already
described.\cite{PostmaTYGD01,JaveyGWLD03,MisewichMATHT03,RueckesKJTCL00,HeinzeTA05}
\\
Yet, for making efficient molecular electronic circuits
multi-probe junctions are also needed and their transport
characteristics must be understood.\cite{YiC03} \textit{Three-terminal
junctions} are especially appropriate for these studies, as they
can be taken as building blocks of multi-terminal junctions. The
understanding of the transport properties of these three-terminal
systems is the goal of this work.
The first experimental observation of three-terminal CNTs were as
branches in L-, T-, and Y-patterns occurring during the growth of
carbon nanotubes produced in an arc-discharge
method,\cite{ZhouS95} but the formation of these junctions was
totally random. Since then, new growth methods have been developed
to obtain these multi-terminal junctions in a more controlled and
high-yield production way: a template-based pyrolytic technique
which yields large numbers of well-aligned Y-junctions of
multi-walled CNTs (MWCNTs, a coaxial arrangement of
CNTs),\cite{LiPX99,SuiGBS01} a hot filament chemical vapor
deposition (CVD) system where uniformly Y-shaped junctions are
obtained as by-product, with structures that match those of
topological models where only heptagons are included as defects in
the hexagonal network,\cite{GanAZYRHYYL00} or pyrolysis of
nickelocene in the presence of thiophene,\cite{SatishkumarTGR00}
or of ferrocene and cobaltocene.\cite{DeepakGR01} Also a simple
thermal catalyst CVD method without templates yields a production
of H- and Y-junctions as well as
3D-CNT-webs.\cite{TingC02,ZhuCXLW02} All these last methods are
characterized by growth temperatures just moderately high
($650\gc$-$1000\gc$) or even quite low (room temperature).
But also the high-temperature arc-discharge technique has been improved
to produce Y-junctions in a reasonable
proportion.\cite{KlusekDBKK02,OsvathKHGBMMB02}
Nearly at the same time, the first observations of three-terminal
single-walled CNTs (SWCNTs) were made where the junctions were
produced by thermal decomposition of fullerenes\cite{NagyEBG00} or
by more sophisticated methods including electron irradiation on
welded nanotubes to finally tailor the transformation of the
junction geometry with the formation of heptagonal and octagonal
defects.\cite{TerronesBGCTA02}

By now, it has been made clear that Y-junctions are not any more a
rare phenomenon. Actually taking into account all the different
methods which yield similar CNT Y-junctions, these have to be
accepted as regular members of the carbon nanostructure family.
Although less experimental data is available, transport
measurements have been carried out on nanotube junctions, by
overlaying one individual SWCNT over another with four electrical
contacts where a good tunnel contact at the junction is
observed,\cite{FuhrerNSFYMCILZM00} or measuring the $I$-$V$
characteristics of truly Y-junction MWCNTs which behave as
intrinsic nonlinear and asymmetric devices, displaying a clear
rectifying behavior.\cite{SatishkumarTGR00,PapadopoulosRLVX00}
Moreover the transport properties of three-terminal junctions 
obtained by merging together SWNTs via molecular linkers 
have also been studied.\cite{ChiuKR04}

From the theoretical point of view, much work has been done to
clarify the structure of these junctions as well as their
intrinsic transport properties. 
To provide an accurate description of quantum transport in CNT-junctions one cannot neglect
their electronic structure via an atomistic model.
Theoretical predictions are then based on hypothesized
atomic configurations,
which try to match the experimental observations on branching angles
and tube diameters. But especially the latter have always a wide
uncertainty.
Most of the experiments on Y-junctions are nevertheless dealing with
MWCNTs and even with a fish-bone like structure as some growth methods
do not achieve a complete graphitization of the tubes. There is
therefore still a need for a better experimental characterization of
the junctions.
Different approaches have been followed to come up with different
proposed structures, like considering the junction as evolved  from
two bend tubes\cite{GanAZHKYRLZL00} or by considering
that three carbon nanotubes join via two triangular central
spacers.\cite{TrebouxLS99}
But no matter how this structure is reached,
non-hexagonal elements must be included in the honeycomb-like lattice,
following the generalized Euler rule,\cite{Crespi98} which 
predicts a bond surplus of six for three-terminal junctions.
These pentagons,
heptagons or octagons are playing an important role in the transport
properties characterizing these junctions, with an electronic
structure which differs considerably from that of the nanotube ``bulk"
region.
Thus several groups\cite{Scuseria92,Chernozatonskii92b,MenonS97,TrebouxLS99,PerezGarridoU02}
have analyzed the geometry of multi-terminal junctions and its stability 
using topological arguments, molecular dynamics techniques or
first-principle methods. The role of the symmetry of the junctions
and of the chirality of the arms is also addressed by many
authors\cite{AndriotisMSC01a,AndriotisMSC02,MeunierNBZC02,MenonASPC03}
as well as the electronic interaction.\cite{ChenTe02} 
However, the controversy remains over the origin of the rectifying
behavior, with some studies claiming it to be a characteristic
intrinsic to the symmetry of the junction,\cite{AndriotisMSC02}
and other works tracing it back to the interfaces with
contacts.\cite{MeunierNBZC02}

Until now Fano resonances have not been exploited as an
interesting feature in the transport through this kind of CNT
junctions. This phenomenon emerges from the coherent interaction
of a discrete state and a continuum and was first discovered by
studying the asymmetric peaks of the helium spectrum.\cite{Fano61}
Recently the occurrence of this effect and its applications have
been studied in numerous mesoscopic devices, including 
MWCNTs\cite{KimKLPSKKYK03,ZhangCDR03} and
MWCNT bundles\cite{YiLHPX03} or SWNT bundles,\cite{BabicS04} and most recently in
multiply-connected CNTs.\cite{KimLKI04} As we will see the
conductance through multi-terminal CNT junctions exhibits Fano
resonances and these may be used in transport allowing for major
changes of the current intensity in short intervals.

\begin{figure}[t]
\centerline{\includegraphics[width=.99\linewidth]{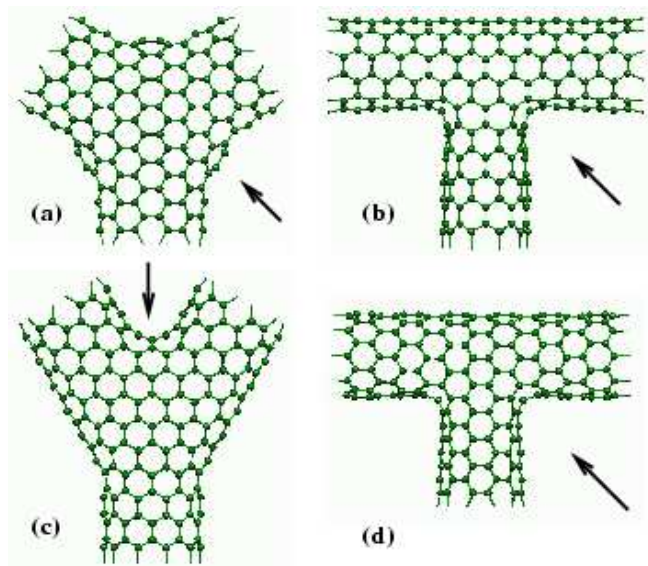}}
\caption{\label{fig:YandTjuncts}
Four junctions entirely made out of semiconducting $(10,0)$ and metallic $(6,6)$ CNT leads. 
(a) Y-junction where three armchair tubes join at $120\g$.  
(b) T-junction where two armchair nanotubes join with a semiconducting
zigzag nanotube. 
(c) Y-junction where three armchair nanotubes join forming an angle of
$72\g$  between the upper arms. 
(d) T-junction where two semiconducting zigzag nanotubes join with an armchair 
nanotube. 
Both Y-junctions have heptagonal rings whereas the T-junctions achieved 
their relaxed structure by the introduction of four octagons and two pentagons. After
relaxation, the average bond length of the heptagons is $2.02$\,\% larger in (a) and 
$1.47$\,\% larger in (c) than the bond length of the hexagonal network. In the case of the 
T-junctions the average bond length of the octagons is in (b) and (d) $1.35$\,\% larger, the bond lengths in
the pentagons not shared with the octagons are nonetheless smaller than that of the 
hexagons by $1.07$\,\% in (b) and $1.31$\,\% in (d). These values are reasonable for having stable
configurations. The defects pointed by the arrows are shown in more
detail in the next figures. A higher definition figure can be found 
in Ref.~\onlinecite{movies}.
}
\end{figure}
The aim of this work is the calculation of transmissions and
conductances for different types of three-terminal CNTs, keeping
in mind the properties of semi-infinite CNTs considered as partial
components of our system analyzed before.\cite{DelValleTC}

 The paper is organized as follows: we will
firstly explain the method used in our calculations and introduce
the different junctions which will focus our attention in this
paper. The presentation and a brief discussion of the results 
for the different transport properties will follow in Sec. III and will 
be concluded in Sec. IV.

\section{System and Method}

Fig.~\ref{fig:YandTjuncts} shows four archetypical three-terminal
CNT devices. These devices have a much greater versatility than
two-terminal junctions as the third terminal can be used to apply
a gate voltage and thus control the current flow in the channel
built by the two other arms, using the gate as a current probe or as a voltage probe. 
Out of the many possibilities of building three-terminal junctions, we have chosen 
these four as to have within a few junctions different elements of comparison: 
different symmetries, different geometries, different chiralities. 
These junctions have nevertheless common properties, from which in conjunction with
their disparities we may draw some general conclusions.
As a model for nanoelectrodes, we will
use semi-infinite carbon nanotubes and try to gain insight in the
quantum transport through these junctions. 
We calculate the quantum conductance within the well-known
Landauer-B\"uttiker formalism\cite{Landauer87,Buettiker88} 
by making use of equilibrium Green
functions for tight-binding Hamiltonians.\cite{CunibertiGG02}
Our structures have been optimized by 
a relaxation algorithm following 
a density-functional based nonorthogonal tight-binding scheme (DFTB)\cite{ElstnerPJEHFSS98} 
in order to deal with stable structures. This method is an extension of the tight-binding formalism,
based on a second-order expansion of the Kohn-Sham total energy in density-functional theory 
with respect to charge density fluctuations. 
Besides the usual  short-range repulsive terms the final approximate Kohn-Sham 
energy additionally includes a Coulomb interaction between charge fluctuations. The basis set used is
minimal but conveniently optimized for carbon atoms. 

The semi-infinite leads are treated with a decimation technique,
an application of the renormalization-group method which allows us
to calculate the electronic properties of extended systems with a
low computational cost. We follow the iterative procedure to
decimate the individual layers\cite{dec} and operate in the space of
localized orbitals.
As principal layers for our system, we take slices
of the CNT and for convenience we choose the CNT unit cell
as unit slice. In this way, we renormalized out 
$2^n$ slices after $n$ iteration steps,
and this exponential behavior allows us to quickly achieve
convergence.

The systems considered are all-carbon devices
for which a $\pi$ orbital description level has been proved to yield very
satisfactory quantitative results,\cite{SaitoDD98} since the properties of
carbon nanotubes are basically determined by $sp^2\,\pi$ orbitals.

\begin{figure}[htbp]
\centerline{\includegraphics[width=.99\linewidth]{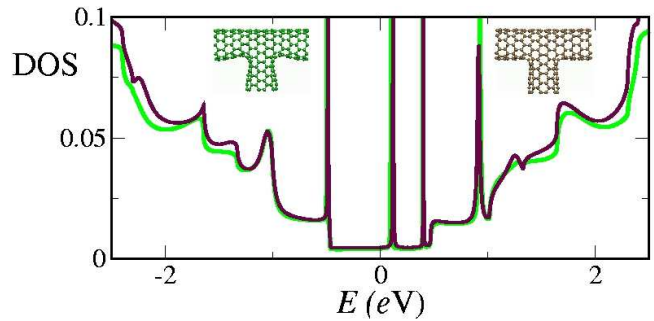}}
\caption{\label{fig:dd}
DOS for the T-junction (Fig.~\ref{fig:YandTjuncts}.d) for the
structure as relaxed using a Tersoff-Brenner potential with
molecular dynamics (lighter color)\cite{PerezGarridoU02} and after
relaxation with DFTB\cite{ElstnerPJEHFSS98} (darker line). Even if
the different relaxation methods yield visibly different
structures at the junction saddle zones, the DOS around the Fermi
energy (with a distance-dependent hopping parameter) is nearly
equal.  A higher definition figure can be found 
in Ref.~\onlinecite{movies}.
}
\end{figure}

The Hamiltonian describing our system can be written in a very compact
form:
\bea
\Ham = \sum_{i,j} h_{ij} \labs i \rab\lab j \rabs
\eea
where $h_{ij}$ is the coupling or hopping parameter.
This transfer integral is only non-zero between
nearest-neighbors and takes the value of $2.66~e$V. We have also made
this parameter distance-dependent, but no significant changes are
observed.
This can be seen in Fig.~\ref{fig:dd}, where the DOS is plotted 
for one of the T-junctions for which another relaxation was available,
making a comparison possible. The DOS calculated with distance-dependent 
hopping parameters for two
differently relaxed structures show an almost identical behavior at low energies.

For calculating the conductance, we divide the system into a central region or
scatterer ($\sr$) 
and the three leads ($\la$, $\lb$, $\lc$).
The Green function for our system, $\Gf\,=\,\lrb \energy\,-\,\Ham \rrb ^{-1}$,
can be written in terms of block matrices
\bea
\lrb\begin{array}{cccc}
\Gf_{\sr}    & \Gf_{\sr\la}  & \Gf_{\sr\lb} & \Gf_{\sr\lc} \\
\Gf_{\la\sr} & \Gf_{\la}     & \Gf_{\la\lb} & \Gf_{\la\lc} \\
\Gf_{\lb\sr} & \Gf_{\lb\la}  & \Gf_{\lb}    & \Gf_{\lb\lc} \\
\Gf_{\lc\sr} & \Gf_{\lc\la}  & \Gf_{\lc\lb} & \Gf_{\lc}
\end{array}\rrb \nonumber \\ =
\lrb\begin{array}{cccc}
\energy - \Ham_{\sr}   & -\Ham_{\sr\la}       & -\Ham_{\sr\lb}       &-\Ham_{\sr\lc}\\
-\Ham_{\sr\la}^\dagger & \energy - \Ham_{\la} & 0                    & 0 \\
-\Ham_{\sr\lb}^\dagger & 0                    & \energy - \Ham_{\lb} & 0 \\
-\Ham_{\sr\lc}^\dagger & 0                    & 0                    & \energy - \Ham_{\lc}
\end{array}\rrb^{-1}
\eea
where the different leads are independent as we restrict our
calculations to nearest-neighbor interaction throughout the whole
system. Note that $\Ham_{{\textrm L}}$ and $\Gf_{{\textrm L}}$ are infinite-dimensional matrices. 
It is straightforward now to obtain via a Dyson equation an explicit expression for
$\Gf_{\sr}$, which has now only finite matrices:
\bea
\Gf_{\sr} = \lrb \energy - \Ham_{\sr} - \Sigma \rrb^{-1}
\eea
where we define $\Sigma =  \Sigma_{\la} + \Sigma_{\lb} + \Sigma_{\lc}$ as the
self-energy terms containing the contribution of the semi-infinite leads
\bea
\Sigma_{\lalle} =  \Ham_{\lalle\sr}^\dagger
\,\Glead_{\lalle}
\,\Ham_{\lalle\sr}
\eea
being $ {\alpha} = 1 $, $2$ or $3$ 
and $\Glead_{\lalle} = \lrb \energy -
\Ham_{\lalle}\rrb ^{-1}$ the lead Green functions.
And these functions are calculated using the decimation technique
mentioned above:
\bea
\Glead = \lrb W_s^{\lrb N\rrb}\rrb^{-1}
\eea
where $N$ is the number of steps necessary to achieve convergence in the iteration process
given by
\bea
W_s^{\lrb i \rrb} &=& W_s^{\lrb i-1 \rrb} - \tau_1^{\lrb i-1 \rrb} \lrb
W_b^{\lrb i-1 \rrb} \rrb^{-1} \tau_2^{\lrb i-1 \rrb} \\
W_b^{\lrb i \rrb} &=& W_b^{\lrb i-1 \rrb} - \tau_1^{\lrb i-1 \rrb} \lrb W_b^{\lrb i-1 \rrb} \rrb^{-1} \tau_2^{\lrb i-1 \rrb}
\nonumber \\ & &
- \tau_2^{\lrb i-1 \rrb} \lrb
W_b^{\lrb i-1 \rrb} \rrb^{-1} \tau_1^{\lrb i-1 \rrb} \\
\tau_1^{\lrb i \rrb} &=& - \tau_1^{\lrb i-1 \rrb} \lrb W_b^{\lrb i-1 \rrb}
\rrb^{-1} \tau_1^{\lrb i-1 \rrb} \\
\tau_2^{\lrb i \rrb} &=& - \tau_2^{\lrb i-1 \rrb} \lrb W_b^{\lrb i-1 \rrb}
\rrb^{-1} \tau_2^{\lrb i-1 \rrb}
\eea
for which we have defined
\bea
W_s^{\lrb 0 \rrb} &=& W_b^{\lrb 0 \rrb} = \energy - H_{{\textrm {PL}}}\\
\tau_1^{\lrb 0 \rrb} &=& - H_{{\textrm {coupl}}}\\
\tau_2^{\lrb 0 \rrb} &=& - H_{{\textrm {coupl}}}^\dagger = \tau_1^{\lrb 0 \rrb \dagger}
\eea
where $H_{{\textrm {PL}}}$ is the Hamiltonian describing any of the principal
layers or slices in which we have divided up the lead, i.e. the projection of 
$\Ham$ onto the space of this slice. These slices are coupled 
to their nearest-neighbor layers in the lead through the interactions
described by $H_{{\textrm {coupl}}}$.

With the lead Green functions we
can easily calculate the
strength of the coupling of the scatterer to the leads 
\bea
\Gamma_{\lalle} \lrb \energy,\mu_{\alpha}\rrb= \ii \lrb \Sigma_{\lalle} \lrb \energy,\mu_{\alpha}\rrb-
\Sigma_{\lalle}^\dagger \lrb \energy,\mu_{\alpha}\rrb\rrb.
\eea
Using all these definitions, we can write the conductance function in
a very compact form:
\bea
\CondEl_{{\alpha\beta}} = \frac{2 e^2}{h} \textrm{Tr} \lcb \Gamma_{\lalle}
\Gf_{\sr}^{\phantom{\dagger}} \Gamma_{\lalleb} \Gf_{\sr}^\dagger \rcb
\eea
where the factor two accounts for the spin degeneracy.
The Landauer-B\"uttiker formula relates the conductance in a
multi-terminal conductor to its
scattering properties. For the three terminal case, if we take one of
the leads to be a voltage probe (say for instance lead $\lc$), we
constrain the chemical potential in this arm to be\cite{Buettiker88}
\bea
\mu_{3} = \frac{\CondEl_{31}\mu_{1}+\CondEl_{32}\mu_{2}}{\CondEl_{31}+\CondEl_{32}}.
\eea
Although this constriction assures a zero current at terminal $\lc$,
this probe is dissipative as the carriers in it loose their phase
memory, accounting for a phase-incoherent contribution to the coherent
current of carriers reaching directly probe $\la$ from $\lb$. The
total conductance can then be written as ~$\Cond = \CondEl + \CondInel$, where
\bea
\CondInel_{12} = \frac{\CondEl_{13}\CondEl_{32}}{\CondEl_{31}+\CondEl_{32}}.
\eea
For our systems the inelastic contribution to the conductance is very
small at energies around the Fermi level.

The current measured in this way is then
\bea
I_{12} = \frac{2 e}{h}\int_{-\infty}^{+\infty}\!\!\!\!\!\!\!\!\textrm{d}\energy\,\lsb f_{\la}\lrb \energy\rrb 
- f_{\lb}\lrb \energy\rrb\rsb \T_{12}\lrb \energy, \mu_1, \mu_2\rrb,
\eea
where $f_{\lalle}\lrb \energy\rrb = \lrb e^{\lrb \energy - \mu_{\alpha}\rrb /k_{\textrm B} {\textrm T}} + 1\rrb^{-1}$ is the Fermi
function in the lead $\lalle$, $\mu_{1}=eV/2$, $\mu_{2}=-eV/2$, and $\T_{12}$ is the total
transmission between lead $\lb$ and $\la$ such that $\Cond_{12}=\lrb 2 e^2/h\rrb \T_{12}$. 
The voltage drop $V$ is small enough to validate the linear-response regime, and we restrict
our calculations to the zero temperature limit.

\section{Transport properties}

We have studied different types of SWCNT junctions, from which four are now presented 
in Fig.~\ref{fig:YandTjuncts},
chosen in such a way that a variety of heterojunction combinations and a variety of 
shapes are covered.  
In these junctions the following groups of three semiinfinite CNTs welded together:
$(6,6)-(6,6)-(6,6)$ in a Y-shape, 
$(6,6)-(10,0)-(6,6)$, both Y- and T-shaped, and $(10,0)-(6,6)-(10,0)$ with a T shape.
Two types of CNT leads have been adopted:  $(6,6)$ armchair tubes and $(10,0)$ zigzag tubes. 
As known $(n,m)$ CNTs, where $(n,m)$ are the indexes unambiguously
determining the characteristics of the CNT, 
are metallic if $n-m$ is a multiple of three, neglecting the $\pi-\sigma$
hybridization. We thus have a metallic tube, the $(6,6)$ CNT and a semiconducting
one, the $(10,0)$ exhibiting a gap of $0.82~e$V.
Then by using just
these two different CNTs we are able to make M-M and M-S
heterojunctions (where M stands for metallic and S for
semiconducting).

\begin{figure}[t]
\centerline{\includegraphics[width=.99\linewidth]{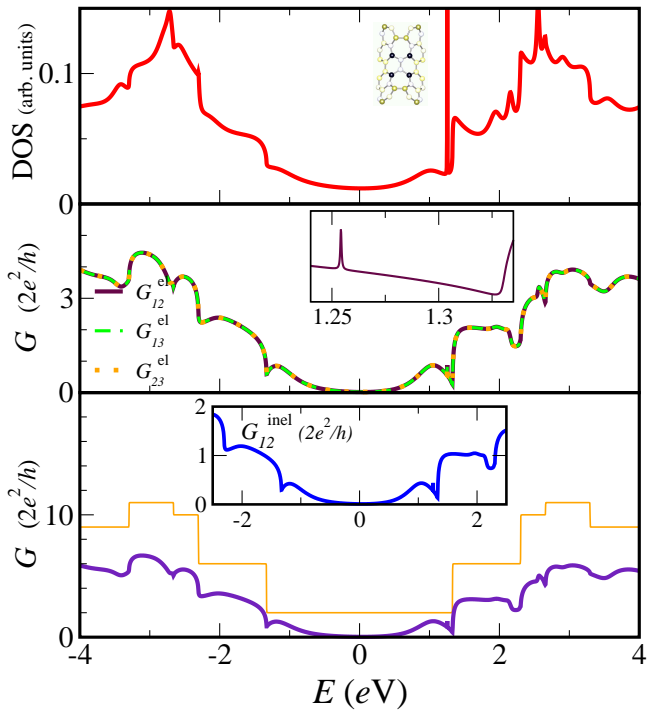}}
\caption{\label{fig:Yjunct}
DOS (projected on the junction) and conductance of the Y-junction
shown in Fig.~\ref{fig:YandTjuncts}.a  
as a function of the incident electron energy. The inset in the upper 
panel shows the defect region pointed by the arrow in 
Fig.~\ref{fig:YandTjuncts}.a with the atomic distribution of the LDOS at 
a localized state, whose conductance can be seen in the inset of the middle panel. 
The LDOS takes values from zero (white) to its maximum value in dark-blue. 
The evolution of the atomic distribution of the LDOS at the junction have been
followed up in an energy window around the Fermi
level.\cite{movies} In the inset of the middle panel we see a
blow-up of $\CondEl_{12}$ near the Fermi energy where resonant
states are observed. The units of the conductance is the
conductance quantum $2e^2/h$. In the lower panel the total
conductance $\Cond_{12}$ for the upper arms of the junction when
making a voltage probe out of the third lead is presented (dark
line) together with the conductance of the perfect infinite CNT of
chirality matching that of the source-drain tubes, in this case the
$(6,6)$ CNT (light line). In the inset here we plot the inelastic
contribution to the conductance through the upper arms. A higher definition figure can be found 
in Ref.~\onlinecite{movies}.
}
\end{figure}

In the chosen configurations, one of the
terminals is taken to be either a voltage probe or a current probe. This gate voltage is
controlled in our model by changing the onsite energies of this arm, which
are fixed at a far end of the terminal and gradually change in the
central region to meet the value set for the other two arms and the
junction.

\begin{figure}[t]
\centerline{\includegraphics[width=.99\linewidth]{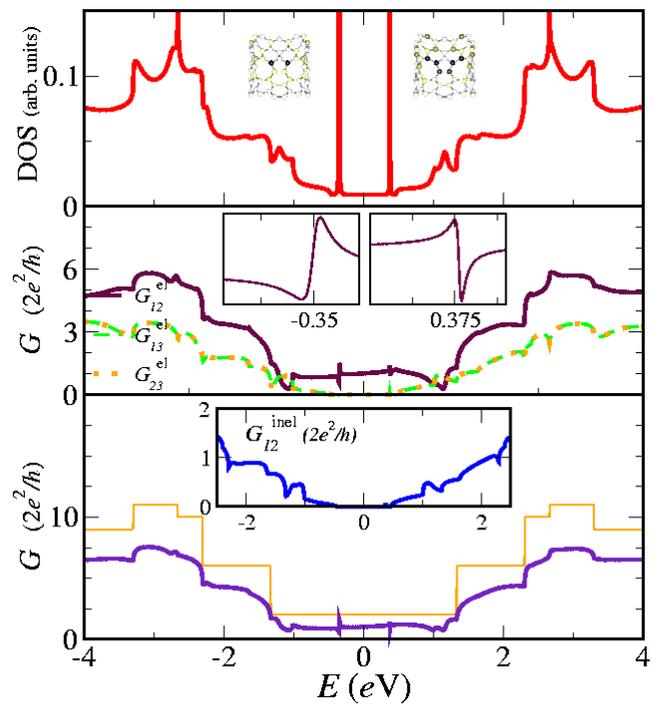}}
\caption{\label{fig:T2junct}
The same properties as in Fig.~\ref{fig:Yjunct} are plotted here for
the T-junction of Fig.~\ref{fig:YandTjuncts}.b. In the insets the
asymmetric line shapes corresponding to a Fano resonance are
clearly observed. A higher definition figure can be found 
in Ref.~\onlinecite{movies}.
}
\end{figure}

\begin{figure}[t]
\centerline{\includegraphics[width=.99\linewidth]{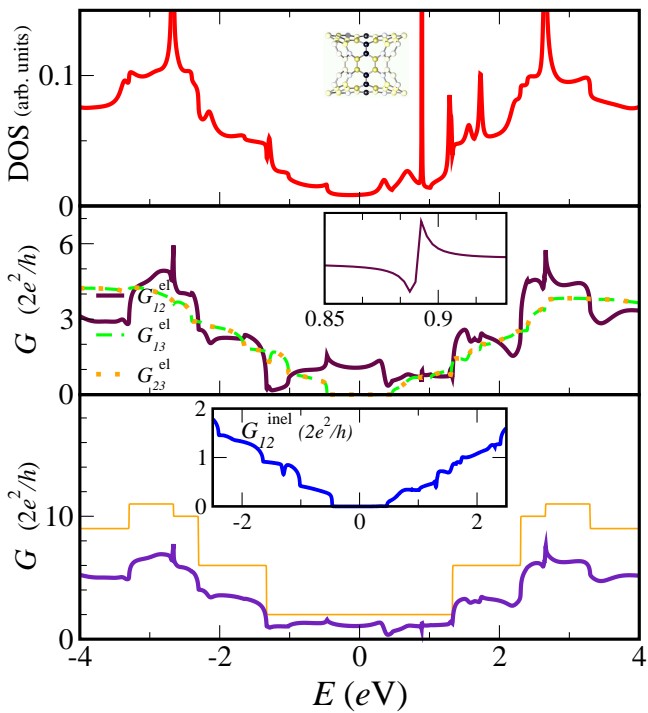}}
\caption{\label{fig:Y2junct}
The same properties as in Fig.~\ref{fig:Yjunct} are plotted here for the Y-junction of 
Fig.~\ref{fig:YandTjuncts}.c. A higher definition figure can be found 
in Ref.~\onlinecite{movies}.
}
\end{figure}

\begin{figure}[t]
\centerline{\includegraphics[width=.99\linewidth]{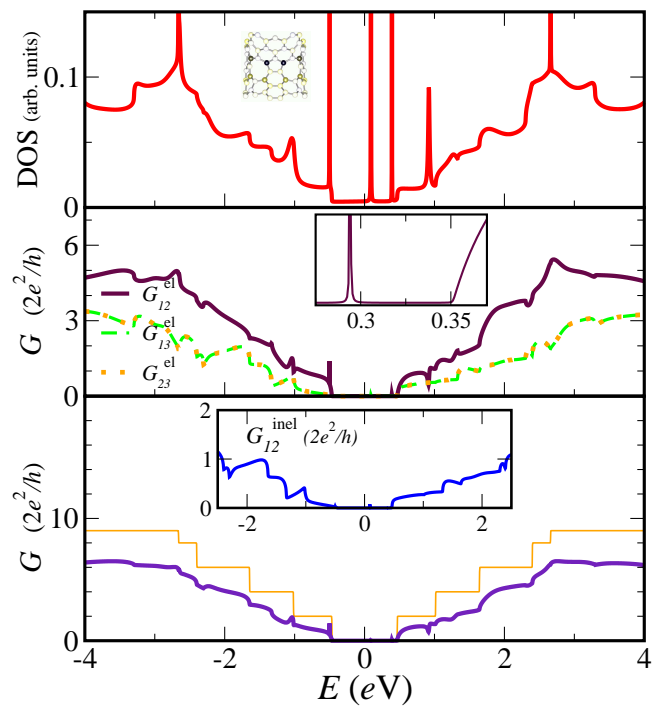}}
\caption{\label{fig:Tjunct}
The same properties as in Fig.~\ref{fig:Yjunct} are plotted here for the T-junction of 
Fig.~\ref{fig:YandTjuncts}.d. A higher definition figure can be found 
in Ref.~\onlinecite{movies}.
}
\end{figure}

We will now proceed to describe the obtained results for the DOS and conductance
of these junctions, which are given in Figs.~\ref{fig:Yjunct}, \ref{fig:T2junct},
 \ref{fig:Y2junct} and \ref{fig:Tjunct}.

Let us consider the electronic structure of the first Y-junction
(Fig.~\ref{fig:YandTjuncts}.a), a highly symmetric junction where
three metallic $(6,6)$ armchair nanotubes intersect at $120\g$. This
structure has mirror symmetry with respect to four planes, being
thus the most symmetrical of all four studied junctions, as the
other three are characterized by two mirror planes. The necessary
negative curvature is provided by the introduction of heptagonal
defects, in a number of six to meet the topological constraints
imposed by Euler theorem, and distributed symmetrically, two at
each saddle region.

This junction shows a metallic behavior, but the transmission
probability around the Fermi level is small as seen in the lower panel of
Fig.~\ref{fig:Yjunct}. The localization of the states at the Fermi energy
is schematically seen in the inset of this figure, showing that this is
an extended state but also pinned by the defects. This is an effect we see along all the
different junctions. It is known that
heptagons and pentagons can induce additional electronic states close
to the Fermi energy,\cite{MenonS97,KlusekDBKK02} which are seen here (upper panel of 
Fig.~\ref{fig:Yjunct}) between the subbands.

The first T-junction (Fig.~\ref{fig:YandTjuncts}.b) is made up of two metallic $(6,6)$ armchair nanotubes
and a semiconducting $(10,0)$ zigzag nanotube. This is achieved by introducing
two octagons and one pentagon at each bending point. This junction also shows a metallic
behavior (upper panel of Fig.~\ref{fig:T2junct}) although one of its branches has a
semiconducting character. Remarkable are the two sharp peaks located
symmetrically around the Fermi level (approximately at $0.35~e$V below
and above it). The distribution of these states in the area of the non-hexagonal defects
is seen in the upper insets of Fig.~\ref{fig:T2junct}.
These are states which do not extend at all to the upper part of the
junction and are mainly localized at the defects. 
To corroborate this thesis, we have checked
the behavior of the local density of states (LDOS) at the resonant energies.  
The presence of both extended states and localized defective states in the junction creates
the conditions for the Fano resonance to be observed.\cite{Fano61}

The resonance-like structure in the transmission exhibits indeed asymmetric line
shapes resembling those of Fano resonances.
These line shapes result from the interaction of a discrete state with a
continuum of metallic states.
The degree of asymmetry of these curves is determined by the so-called Fano
parameter, which changes here its sign.
Therefore at the resonances we have that
$\T\propto \lrb q+E \rrb^2/\lrb 1+E^2\rrb$ where
$q= \ReP \Gf^0/\ImP \Gf^0$ is the Fano parameter
and $\Gf^0$ is the undressed Green function for the continuum.
Due to the analyticity properties of the Green functions (causality), the real
and imaginary parts of $\Gf^0$ are related by a Kramers-Kronig (Hilbert)
transformation:
\bea
\ReP \Gf^0 \lrb \energy
\rrb=\frac{1}{\pi}P\int_{-\infty}^{+\infty}\frac{\ImP \Gf^0\lrb
\energy'\rrb}{\energy'-\energy}\,\textrm{d}\energy',
\eea
where $P$ means principal value.
From these equations, we see that for an asymmetric DOS with most
of its weight situated at frequencies below the energy
$E_{\textrm{\,Fano}}$ in a small window around this energy, it is
likely to obtain $q>0$, whereas for asymmetric DOS with more
weight at $\energy>E_{\textrm{\,Fano}}$, $q$ is likely negative.
The shape of the DOS of the continuum of states of this metallic
junction decreases rapidly before the left peak, and increases
just after the one to the right. This accounts for the different
signs of $q$ as observed in the plot. Though the states around
$E_{\textrm{\,Fano}}$ are the most significant for determining the sign of $q$, it is necessary to
take into account the whole range of energy of the band.

\begin{figure}[t]
\centerline{\includegraphics[width=.99\linewidth]{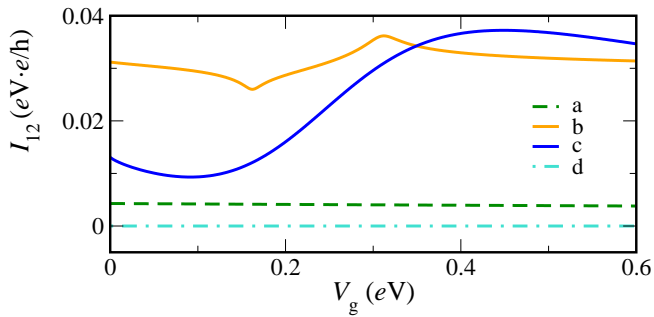}}
\caption{\label{fig:gate}
%
Current vs.~gate voltage using one arm of the three-terminal junctions as gate
electrode ($\lc$). The switch-on of the gate shows to be of interest for possible
future applications of these junctions, and especially of the
second of the Y-shaped ones. The labels a, b, c, d correspond to the notation given in 
Fig.~\ref{fig:YandTjuncts} for the four junctions considered. A higher definition figure can be found 
in Ref.~\onlinecite{movies}.
}
\end{figure}

These characteristics are then observed in the $I$-$V$ curves
(Fig.~\ref{fig:gate}.b), though its effect is not that of a
switch-off of the current, as the transmission background of the
continuum is too strong. Nevertheless differences in the current
amplitude up to 30\% are observed.

The second Y-junction, shown in Fig.~\ref{fig:YandTjuncts}.c, is made up with the same combination of
leads as the previous T-junction, but forcing now the armchair tubes to bend, forming a
Y-shape. In this case, we have a very different rearrangement of the defects. 
Four of the heptagons are situated on the upper saddle
while the other two are shared by the obtuse angles. The junction is indeed metallic, 
as shown in Fig.~\ref{fig:Y2junct}, but as in
the previous case quickly assumes the electronic character of the corresponding arms. 
Above the Fermi energy, the necessary conditions for a Fano resonance appear again. The sharp 
peak of a localized state is seen in the DOS and its corresponding Fano resonance 
in the conductance. Its behavior all in all resembles that of the
junction in Fig.~\ref{fig:YandTjuncts}.b but anyhow presents remarkable differences, as a
greater asymmetry, which are exclusively due to the junction shape and the 
situation of the defects.

The $I$-$V$ characteristics of this junction shown in Fig.~\ref{fig:gate}.c present an
interesting profile, as the big change in current makes it interesting for its use as a circuit
component.

The last of the presented junctions shown in Fig.~\ref{fig:YandTjuncts}.d is again a T-shaped one 
but built up of one metallic and two semiconducting tubes. Like in the other T junction 
the defects are octagons and pentagons,
distributed symmetrically in both bending regions. The transport properties of this junction,
given in Fig.~\ref{fig:Tjunct}, are dominated by the semiconducting behavior of the upper branches,
whose gap is reflected in the conductance of this junction. A non-equilibrium study of its
properties would be necessary to fully explore its possible applications in electronics.

The influence of the gate voltage on the current is shown in Fig.~\ref{fig:gate}. Here the
current flowing between the upper branches is calculated under a small bias voltage of the order of a
few tens of m$e$V, while we smoothly change a gate voltage which shifts the
levels of the orbitals at the third arm of the junction. We should
stress at this point that the bias voltages used in this calculation allows us to remain within the
linear response regime. As can be
seen the current suffers a great modulation in some of the junctions, increasing its amplitude in up to
75\%.

\begin{figure}[htbp]
\centerline{\includegraphics[width=.99\linewidth]{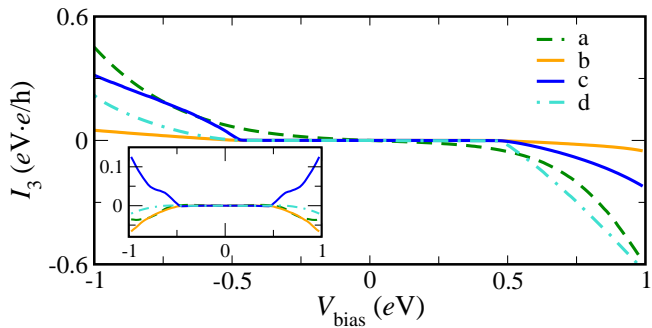}}
\caption{\label{fig:CurrentAndriotis}
Current $I_3$ vs. bias voltage for the four junctions of
Fig.~\ref{fig:YandTjuncts}. $I_3$ is the total current through
lead $L_3$, that is $I_3 = I_{31} +I_{32}$. The current is calculated for 
the experimental setup described in Eq.~\ref{setup1} as well as for a 
different situation given by Eq.~\ref{setup2} and plotted in the inset.
As in the previous figure the 
labels a, b, c, d correspond to the notation given in 
Fig.~\ref{fig:YandTjuncts} for the four junctions we consider. A higher definition figure can be found 
in Ref.~\onlinecite{movies}.
} 
\end{figure}

Despite the fact that our model is adequate for getting linear response, 
we apply it to a nonlinear situation for
sheer comparison of our results to previous calculations.\cite{AndriotisMSC02}
We calculate the $I$-$V$ characteristics of these junctions
considering different possible experimental setups. In
Fig.~\ref{fig:CurrentAndriotis} we report the current through the third lead in two different situations: 
first the upper branches are grounded and the third lead is
biased with a voltage $V_{{\textrm {bias}}}$. That is, 
\bea 
\label{setup1}
\mu_1=\mu_2=0;\, \mu_3=V_{{\textrm {bias}}}. 
\eea

Under these conditions a certain degree of rectification power is
present in the setup, but the rectification is not an inherent
property of the junctions. This can be seen in a second case where the third 
lead is grounded while a bias is applied between the first and second lead, i.e.
when plotting the
current versus the bias voltage for the setup

\bea 
\label{setup2}
\mu_1=-\mu_2=-V_{{\textrm {bias}}}/2;\, \mu_3=0, 
\eea

as shown in the inset of the same figure. The degree of asymmetry in view at the 
$I$-$V$ curves when using the first of the experimental setups disappears for the second setup. 
In this case the current curves are symmetrical with respect to
the bias voltage, both for $I_3$, plotted in the inset, as for the current flow through 
the upper arms.

\section{Discussion and conclusions}

By making use of the Landauer formalism and equilibrium Green
functions techniques, we have studied the transport properties of
different carbon nanotube junctions, at an atomistic description
level, and paid especial attention to bound states. Our analysis
of the conductance behavior of several three terminal carbon
nanotube junctions allows us to identify common patterns and draw
general conclusions. This can be done by sorting out different
factors in the qualitative features of the conductance, namely (i)
the role of structural defects in the molecular network, (ii) the
resulting geometrical symmetry properties, and (iii) the
electronic nature of the contacted leads (whether metallic or
semiconducting tubes). By opening the energy of the electron
incoming in the injecting lead one encounters typical resonant
behaviors of the LDOS.\cite{movies} For such energies most
spectral power is associated to the defective atoms at the saddle
regions. The presence of background states, mostly contributed by
the attached leads, creates the conditions for an interference
between localized and extended states, with the appearance of the
so-called Fano resonance. Indeed resonant defective states in the
DOS are paired by the typical Fano line shape of the form 
\bea
\T\propto \frac{\lrb q+E \rrb^2}{1+E^2}. 
\eea
This feature could be eventually detectable in experiments, as 
differences in current amplitude up to 75\% are observed when using 
one arm of the three-terminal as a gate electrode. We have thus shown 
how through the localized states, we can control the current flow through 
the upper branches which is driven by a bias voltage applied across
the first two terminals. 


\begin{acknowledgments}
We are indebted to Marieta Gheorghe and Pasquale Pavone  
for helpful support at different stages of this work,
and to Antonio P\'{e}rez Garrido for providing us with the coordinates of one of the junctions.
MD acknowledges the support from the FPI Program of the Comunidad Aut\'{o}noma de Madrid.
This work was funded by the Volkswagen Foundation (Germany), by the MCYT (Spain)
under the contract MAT2002-00139, by the EU within the RTN COLLECT and by the EU-FET program under
contract IST-2001-38951.
\end{acknowledgments}



\begin{thebibliography}{57}
\bibitem[{dec()}]{dec}
\bibinfo{note}{The decimation scheme, outlined in
  Ref.~\onlinecite{GuineaTFL83}, proved to us to meet highest computational
  accuracy. A similar algorithm has been proposed in
  Ref.~\onlinecite{LopezSanchoLSR84} and extensively applied to CNT
  calculations, see for instance Ref.~\onlinecite{NardelliSep99}.}
\end{thebibliography}
\end{document}